\documentclass{svjour3}
\usepackage{graphicx}
\usepackage{url}
\usepackage{xspace}
\usepackage{amsfonts}
\usepackage{color}
\def\systemname#1{\textsf{#1}\xspace}
\def\libname#1{\textsf{#1}\xspace}

\newcommand{\mizar}{\systemname{Mizar}}
\newcommand{\Mizar}{\systemname{Mizar}}
\newcommand{\MizAR}{\systemname{Miz$\mathbb{AR}$}}
\newcommand{\mml}{\libname{MML}}
\newcommand{\MML}{\libname{MML}}
\newcommand{\SOTPTP}{\systemname{SystemOnTPTP}}

\newcommand{\mizby}{`{\tt by}'\xspace}
\newcommand{\mizbys}{`{\tt by;}'\xspace}
\newcommand{\mizfrom}{`{\tt from}'\xspace}

\title{ ATP and Presentation Service for Mizar Formalizations}
\author{Josef Urban \and Piotr~Rudnicki \and Geoff~Sutcliffe }

\institute{  Josef Urban, funded by 
NWO grants \textit{MathWiki} and \textit{Knowledge-based Automated Reasoning} \at  Radboud University, Nijmegen \and Piotr~Rudnicki, supported by a NSERC
grant \at University of Alberta \and Geoff~Sutcliffe \at University of Miami}

\begin{document}
\maketitle

\begin{abstract}

  This paper describes the {\em Automated Reasoning for Mizar} (\MizAR)
  service, which integrates several automated reasoning, artificial
  intelligence, and presentation tools with \Mizar and its authoring
  environment. The service provides ATP assistance to \Mizar authors
  in finding and explaining proofs, and offers generation of \Mizar
  problems as challenges to ATP systems. The service is based on a sound
  translation from the \mizar language to that of first-order ATP
  systems, and relies on the recent progress in application of ATP systems in large
  theories containing tens of
  thousands of available facts. We present the main features of \MizAR
  services, followed by an account of initial experiments in finding proofs
  with the ATP assistance.  Our initial experience indicates
  that the tool offers substantial help in exploring
  the \mizar library and in preparing new \mizar articles.

\end{abstract}

\vspace{-1em}
 \section{Motivation and System Overview}

Computer supported formal mathematics~\cite{Hal08}
is becoming better known, widely used and experimented with.
Projects like FlySpeck \cite{Hal05}
and verification of tiny (but real) operating systems 
\cite{KE+09}
are stimulating the development of interactive
verification tools and interactive theorem provers (ITPs).
Linked to this is the development of strong automated theorem proving (ATP)
systems, used either independently to solve hard problems in suitable domains,
or integrated with interactive tools.
Such integrations have motivated recent  research in the context of
automated reasoning in large theories
\cite{MengP09,PS07-ESARLT,cade/UrbanSPV08}.

The \mizar project\footnote{\url{http://mizar.org} . See~\cite{GKN10}
  for introductory information on \mizar. For the rest of the paper
  we assume at least superficial familiarity with \mizar.} is a long-term
effort to collaboratively
develop a formal computerized system representing important mathematical knowledge. The project is focused on building the \mizar
Mathematical Library (\mml) -- a collection of mathematical `articles'
formalized and mechanically verified with the \mizar{} system.  \mml
is the largest such library among similar projects.

This paper introduces the {\em Automated Reasoning for Mizar} (\MizAR)
service, which attaches several automated reasoning and presentation
tools to the \mizar system.
\MizAR runs in the context of the \MML 
expressed in the \Mizar language -- a language oriented
toward mathematicians.  An inspiration for \MizAR is the
established {\sf SystemOnTPTP} ATP service \cite{Sut00-CADE-17}, which uses
the simpler TPTP language for expressing proving tasks.
\SOTPTP provides a framework 
for finding proofs with many first-order ATP systems, offers
various forms of proof presentation,
supports discovery of new lemmas 
and independent proof verification.
The \MizAR service links to and re-uses parts of \SOTPTP.
\MizAR comprises the following main functionalities:
\begin{itemize}
\item Fast server-based  verification of
  \Mizar articles, and their HTML-based disambiguation linked to the whole cross-linked HTMLized \MML -- see Section~\ref{sect:server}.
\item Generation of ATP problems in TPTP format for theorems and inferences
  in a given article, and invoking automated provers on them -- see
  Sections~\ref{TPTP}, \ref{UsingWholeMML}.
\item Searching for useful premises for \Mizar lemmas and
  inference steps -- see Sections~\ref{sect:fromEmacs}, \ref{GettingHints}.
\end{itemize}
The first version of the
\MizAR service was deployed in 2008 on a server of the Automated Reasoning 
Group in Prague. 
As of 2011 the service is independently hosted on two mid-range multi-core 
servers in Nijmegen\footnote{%
  \url{http://mws.cs.ru.nl/~mptp/MizAR.html}} and in Edmonton\footnote{%
  \url{http://mizar.cs.ualberta.ca/~mptp/MizAR.html}}.
A way for newcomers to explore \MizAR is to use a web browser
with an existing simple \mml article, e.g., the {\tt CARD\_1}
article\footnote{%
  \url{http://mws.cs.ru.nl/~mptp/mml/mml/card_1.miz}} about cardinal
numbers \cite{Ban90} from the \MML.
More experienced \Mizar users will typically interact with the service
by launching commands from the \Mizar Emacs authoring
interface~\cite{Urb06-JAL}. Some of the commands keep the
communication fully in the Emacs session, while other commands
offer a browser-based interaction.
\section{Server-based Verification and HTMLization}
\label{sect:server}

The \Mizar verifier acts as a compiler-like batch processor, verifying an entire
article in one pass and reporting errors.
The process of checking a whole article can be quite time-consuming
for longer and more complex articles. %
In the omni-presence of the fast internet, there are several advantages
of remote server-based verification of \Mizar articles:
(i) no need for a local installation of the entire system; (ii) 
browser-based access for simple tasks; (iii) large numbers of faster CPUs on
the servers, offering great speedups through parallelization~\cite{abs-1206-0141}.
A dedicated server-based installation can also support modified,
enhanced, or experimental versions of the verifier.
For instance, an enhanced version of the verifier is useful when
translating \Mizar to ATP formats, and the \Mizar parallelizer
requires some Linux tools that might not be available on other
platforms.\footnote{As of 2011, \Mizar is
  distributed for eight architectures, some of them targeted at PDAs.}
An online service can also easily include multiple versions of the \Mizar 
library. %
This opens a path toward a wiki-like collaborative environment
for formalizing mathematics in \Mizar
\cite{UrbanARG10}.

\Mizar articles are written as text files according to the \mizar
syntax, but their semantics is defined by the verifier in the context
of the \mml.  Discovering all semantic details of such a formal article by
hand from its raw textual form can be a challenging task, because of
overloaded notation that is so common in mathematical practice.  Fortunately,
the internal format of \mml is XML based~\cite{Urb05-MKM}, which can be
automatically turned into an annotated HTML presentation.  The HTML
presentation, while close to the original text of the \mizar article,
offers assistance in semantic browsing both through the \mml as well
as displaying semantics of an article still in development.  Such assistance includes displaying the current goal (thesis)
computed by the verifier at each point of a proof, disambiguation of
overloaded mathematical symbols through hyperlinks,
and explicit display of formulae hidden 
behind certain keywords 
announcing properties (such as projectivity, commutativity,
antisymmetry, etc.) of constructors.  The HTML presentation of \mizar
texts forms the backbone to which other \MizAR services are
linked.
\section{Translation to ATP Formats and Integrating ATP systems}
\label{TPTP}

\MizAR provides access to ATP systems in  the context of the large
body of mathematics in the \MML. The library is first translated into the
MPTP (\Mizar Problems for Theorem Provers) intermediate
format~\cite{Urban06}, and then translated into the TPTP format that is 
the standard for many ATP systems.  
Complete static versions of the \MML in the MPTP and TPTP
formats are also stored on the \MizAR server, and used for on-demand
conversion of \MML items (theorems, definitions, formulae encoding
implicit \Mizar type, etc.) into ATP problems in various usage
scenarios. The conversion of \mml items into MPTP and then into TPTP
format requires a quite complex installation and setup (including SWI
Prolog, Unix utilities, special XSL style sheets, the \MML
in the MPTP format, etc.) and therefore is better suited for
processing on a dedicated server.

The HTMLization of an article and the generation of ATP
problems are independent processes, and they constitute separate
services that can run in parallel in different CPUs.
For example, a call for ATP help issued from the Emacs
interface would trigger only fast ATP processing,
responding directly to Emacs. Notation disambiguation and proof
explanation tasks would typically also trigger HTML processing,
possibly linking additional ATP and explanation services (running on
the translated article) to the HTML presentation.

The first version of \MizAR used the E and SPASS provers by default,
with an option to export the generated problems to the dozens of ATP
systems and model finders available through the \SOTPTP interface.
In 2010, the default ATP was changed to Vampire, motivated by its
improved behavior on \Mizar/MPTP problems, by its general
compliance with the TPTP format, and
particularly because of the direct integration of the SInE
premise selection method 
\cite{HoderV11}.\footnote{
This changes quickly: a fast SInE algorithm has been recently added also to the E prover.} 
A recent comparison~\cite{icms/UrbanHV10} of Vampire with
the E and SPASS ATP systems on the set of all theorems from MML
version 1011 is summarized in Table~\ref{Repr2}. This comparison is
based on the \emph{SMALL} versions of the \MML problems, 
in which only the premises explicitly provided by \Mizar authors 
(plus some general implicit background facts) 
are used for constructing the problems. In this mode Vampire
solves 20109 problems out of 51424 within a 30s
CPU time limit per problem, which is significantly better than (untuned)
E, solving 16191 of the problems. For more details and statistics of other usage scenarios, see~\cite{icms/UrbanHV10}.
The ATP systems are typically run (possibly in parallel) with
different premise selections (even if some of them do their own
premise selection internally), depending on the usage scenario. Some
of the scenarios are explained below. 
\begin{table*}[htbp]
  \caption{Evaluation of E, SPASS, and Vampire on all \MML \emph{SMALL} problems in 30s}
  \begin{tabular}{|l|r|r|r|r|r|}
    \hline
    description&proved&countersatisfiable&timeout or memory out&total\\
    \hline
    E 1.1-004 &16191&4&35229&51424\\
    \hline    
    SPASS 3.7&17550&12&33862&51424\\
    \hline
    Vampire 0.6&20109&0&31315&51424\\
    \hline
    together &22607&12&28817&51424\\
    \hline    
  \end{tabular}
\label{Repr2}
\end{table*}

\vspace{-2em}

\section{Solving Problems with the Use of the Whole MML}
\label{UsingWholeMML}
An obvious use-case of \MizAR is when a new conjecture is attacked
with the help of the whole \MML library, containing about a hundred
thousand premises.  While there are several complementary AI
approaches to premise selection, and experimenting with them is 
interesting and potentially very rewarding, the default 
method for this use-case is the \Mizar-tuned Vampire/SInE 
system, which is capable of loading the whole translated \mml and
selecting promising premises in seconds.

When a user asks the service to solve a problem
  using the whole \MML, the service creates a TPTP problem for the task 
  by including the file containing the whole translated MML
  (available statically on the server in TPTP format), and
  adding 
  all the propositions from the current article that are available 
  before the proposition for which a proof is sought. 
Other
  (typically leaner) premise selections can be created in parallel by
  analogous mechanisms, producing several versions of the problem that
  are handed over to the ATP systems in parallel. The current 
  implementation uses four different premise selections: (i)
  using the full library, (ii) using only premises from the articles
  imported by the current one, (iii) using only premises from the
  current article, and (iv) using only the premises explicitly given
  by the user for the problem. 
As noted above, the current choice is
  to use only Vampire/SInE, and parallelize with
  respect to the different premise selections. This is rather 
  accidental: arbitrary (parallel) combinations of ATP systems and premise
  selection methods are possible, and limited only by the time limit
  and the number of free CPUs on the server.

As soon as a proof is found by at least one of the methods in the
current pool, the TPTP output is searched
  for the necessary axioms, and they are presented to the user either in
  HTML or in Emacs (see below).

\section{ATP-supported Authoring in Emacs}
\label{sect:fromEmacs}
\label{sect:access}

Even though \MizAR is a web-based service in the spirit of \SOTPTP,
this does not mean that it requires a browser to use.  The above mentioned \emph{whole-library solving} functionality is
available directly from the Emacs authoring environment for
\mizar~\cite{Urb06-JAL}, providing fast authoring support without any need for switching to
a browser. The implementation uses the Emacs Lisp {\tt url} module
and a {\tt http-post} request sent directly to the \MizAR
server.  This communication channel also allows other remote
functions, in particular it is possible to call \MizAR only for
remote (parallelized) verification using the raw
speed of the server.

A basic use of \MizAR is illustrated by the following example.
Inference steps are presented to the
\mizar-verifier by stating the goal followed by the keyword \mizby 
with a list of premises. For example,

\begin{small}
\begin{verbatim}
   A: x in L ...
      ...  
   D: {x} c= L by A, ZFMISC_1:37;
\end{verbatim}
\end{small}
where the label {\tt ZFMISC\_1:37} refers to a fact imported from
\mml.\footnote{See \url{http://mizar.uwb.edu.pl/version/7.11.07_4.160.1126/html/zfmisc_1.html#T37}
for the exact statement of {\tt ZFMISC\_1:37} in \MML version
7.11.07\_4.160.1126 . For other theorems cited in this paper, replace
the article name and theorem number accordingly. For definitions,
replace `{\tt T}' by `{\tt D}'.} 
Finding the necessary references requires detailed knowledge of \mml,
and in more complicated cases it is a time-consuming process. 
With
\MizAR available, is is possible to try find sufficient premises (like {\tt
  ZFMISC\_1:37}) by invoking the service with typing \mizbys after
the goal for which assistance is desired.
The query is posted to the \MizAR server while the Emacs buffer changes to

\begin{small}
\begin{verbatim}
   A: x in L ...
      ... 
   D: {x} c= L ; :: ATP asked ... 
\end{verbatim}
\end{small}
This communication is asynchronous, allowing multiple queries.
The ATP answer is provided within seconds, depending on
preset time limits and the server's load.  The premises used in the ATP
solution are used to replace the original `\mizbys' (or a failure is reported).
In this example the result is

\begin{small}
\begin{verbatim}
   A: x in L ...
      ... 
   D: {x} c= L by A,ENUMSET1:69,ZFMISC_1:37;
\end{verbatim}
\end{small}
after which (still in Emacs) the standard \mizar utility {\sf
  relprem} that detects unnecessary premises in an inference can be invoked. 
In this example {\sf relprem} detects that
{\tt ENUMSET1:69} is unnecessary, and
its removal yields the inference step that started this example.

The above example is an inference step taken from a proof of a
very simple theorem in the {\tt SCMYCIEL} article~\cite{RudnickiS12} 

\begin{small}
\begin{verbatim}
   theorem Sub3:
   for G being SimpleGraph, L being set, x being set
    st x in L & x in Vertices G
     holds x in Vertices (G SubgraphInducedBy L) 
\end{verbatim}
\end{small}
Proving the whole theorem is too hard for the ATP service
at the moment, and calling \MizAR fails with the message:
{\tt Sub3: ... Unsolved}.
As an alternative, the following obvious intermediate proof steps can
be tried.

\begin{small}
\begin{verbatim}
   proof
    let G be SimpleGraph, L be set, x be set such that
   A: x in L and
   B: x in Vertices G;
   C: {x} in G ;
   D: {x} c= L ;
   E: {x} in (G SubgraphInducedBy L) ;
     thus x in Vertices (G SubgraphInducedBy L) ;
   end;
\end{verbatim}
\end{small}
None of the sentences labeled {\tt C}, {\tt D}, {\tt E} or the proof
conclusion are obvious to \mizar.  Again, ATP systems can be used by 
replacing semicolons with \mizbys.  The replies from ATP
come almost immediately, with the final result as follows.

\begin{small}
\begin{verbatim}
   C: {x} in G by B,SCMYCIEL:5;
   D: {x} c= L by A,ENUMSET1:69,ZFMISC_1:37;
   E: {x} in (G SubgraphInducedBy L) by C,D,BOOLE:7,SCMYCIEL:14;
     thus x in Vertices (G SubgraphInducedBy L) by SCMYCIEL:func 5,E,BOOLE:7,SCMYCIEL:5;
\end{verbatim}
\end{small}
Such replies from the ATP often
need some post-editing to satisfy the \mizar checker:
\begin{itemize}
\item Some references returned by the ATP service, like {\tt SCMYCIEL:func 5},
  mention typing items, which are implicit\footnote{The
    rich \mizar type system becomes explicit when translated to
    untyped first-order logic.}
    to \mizar and cannot be explicitly referred to.
  \item Some references, like {\tt BOOLE:7}, encode \mizar automations
    (called \emph{requirements}), i.e., theorems added automatically to proof search by \mizar. They typically do not have to be used explicitly
    in \mizar because their automated use is switched on by a global directive.
  \item Some references are spurious for the \mizar verifier (caused
    by the non-minimized ATP proof search) and they can be removed,
    e.g. {\tt ENUMSET1:69} above. Some reference minimization can
    be done with the {\sf relprem} utility.
\item The article is named {\tt SCMYCIEL}, and
  references to lemmas from this article use this name.  These
  references have to be renamed to the corresponding local names.
\end{itemize} 
Most of this post-editing can be automated. Note that studying the
references found by the ATP is instructive as the automated service
sometimes (particularly with a large library) finds solutions quite
different from what the author had in mind. After the post-editing, the
final result accepted by the \mizar verifier is:
\begin{verbatim}
   C: {x} in G by B,Vertices0;
   D: {x} c= L by A,ZFMISC_1:37;
   E: {x} in (G SubgraphInducedBy L) by C,D;
     thus x in Vertices (G SubgraphInducedBy L) by E,Vertices0;
\end{verbatim}
An initial evaluation of this authoring assistance is provided below,
in Section~\ref{sect:experience}.

\section{Access from HTML}
\label{ATPCalls}
\label{sect:essence}
\label{GettingHints}

ATP and other services can be  called from
the \MizAR web interface, which is similar to that of \SOTPTP.
The services can be invoked also from the 
HTML presentation of the user's article, created either through the web
interface or by launching a web browser directly from Emacs.
The HTML presentation contains links to the ATP services that are associated with the
\Mizar keywords \mizby and \mizfrom, indicating logical
justification in \Mizar.  Consider for example, the \Mizar justification

\begin{small}
\begin{verbatim}
    hence ( f is one-to-one & proj1 f = X & proj2 f = A ) 
      by A1, A2, WELLORD1:def 7, WELLORD2:16, WELLORD2:def 1;
\end{verbatim}
\end{small}
in the last line of the proof of theorem {\tt Th4} in the {\tt
  CARD\_1} article. Such justifications may involve many
implicit \Mizar facts and mechanisms that
make the raw \mizar text hard to understand.  The process of
translation to TPTP reveals all this implicit information and the
ATP proofs can show explicitly how this information is used.  For
the \Mizar justification above, clicking on the \mizby keyword
calls the default ATP system
on the corresponding ATP problem.
If a
proof is found, the interface is refreshed with an explanation box
that includes a list of the
references used in the proof. %
In
this case the references shown to the user are 

\begin{small}
\begin{verbatim}
   dt_c2_6__mtest_1, dt_k2_wellord1, dt_k1_wellord2, dt_c5_6__mtest_1, 
   e7_6__mtest_1, e2_6__mtest_1, t16_wellord2, d1_wellord2, e8_6__mtest_1, 
   e6_6__mtest_1, d7_wellord1, 
\end{verbatim}
\end{small}
These references are reported using the MPTP syntax and are linked
dynamically to the corresponding places in the article's HTML or in the
HTML-ized \MML.  Note that the ATP proof reports more references
than in the original \Mizar inference. The extra references are mainly typing
statements used implicitly by \Mizar.

A byproduct of this ATP explanation feature is the cross-verification of
\Mizar atomic inferences.  With a recent version of MPTP and the 
strong ATP systems available at the time, over 99\% of \mizar atomic 
inferences
could be cross-verified~\cite{mics/UrbanS08}.  Such functionality
is valuable as a debugging tool for \mizar developers, and also for
the developers of the MPTP translation layer.

Another interactive mode of use is for generating problems and finding
proofs that are too hard for the \Mizar checker, and experimenting
with the ATP strength in the mathematician-oriented \Mizar language instead of
having to encode the problems directly in the low-level TPTP language.
Users can do this within \MizAR by providing a set of 
premises on the right-hand side of
the \mizby keyword
and letting ATP systems try to find a proof. 
If the default ATP systems are not successful, the user can use the links and
icons in the explanation box to inspect the ATP problem, and
launch the {\sf SystemOnTPTP} interface to try the ATP
systems available there.  In a similar way, one can use ATP systems
and model finders for detecting countersatisfiability of
\Mizar-formulated problems.

It is hard to enumerate all the 
ITP-ATP use-cases that are possible through \MizAR.
For instance, the user might prefer to use a
SAT solver (for attacking propositional problems),
instantiation-based systems like iProver %
(strong in effectively propositional problems), or
to experiment with SMT solvers. 
A reliable ITP-to-ATP translation saves the developers of
ITPs a large amount of work by allowing them to be always on top
of the state-of-the-art in ATP research. Similarly,
a link from the ITP user 
interface to an ATP user interface saves the developers of ITP user
interfaces (in this case the first author) years of work done by
the developers of ATP user interfaces.
While a basic direct implementation makes sense in both of these
cases,
fully reimplementing every new ATP method (or user
interface to it) inside ITPs and their interfaces 
can hardly catch up with the rapid development of
ATP systems and their interfaces.

\begin{figure}[h]
\begin{center}
    \includegraphics[width=0.8\textwidth]{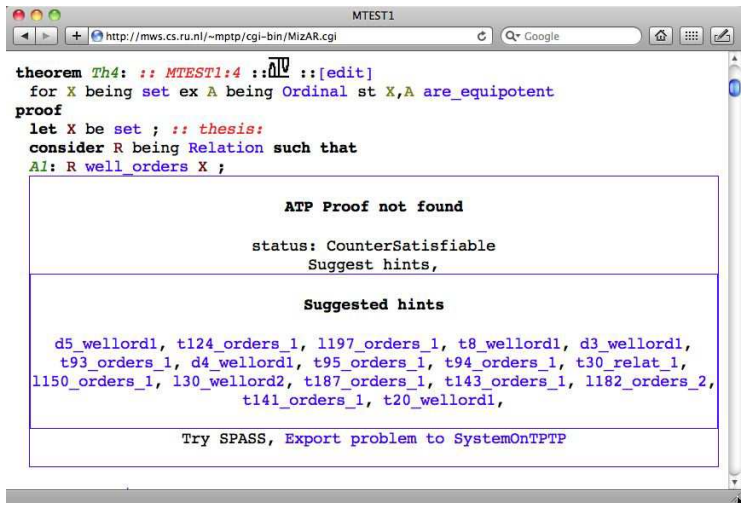}
  \caption{Explanation box offering hints}
  \label{Hints}
\end{center}
\end{figure}

A special kind of service that is particular to ITPs with
large libraries is premise selection based on (possibly
expensive) AI-based preprocessing of the libraries.
If no ATP system can find a proof for a \MizAR-generated ATP problem,
finding relevant premises from \MML can help.  
When ATP fails to find a proof, the `{\tt Suggest hints}' link can be used
to ask \MizAR to suggest a set of potential premises.
This invokes a 
Bayesian advisor that has been trained on the whole \MML (i.e., on all
of the proofs in it).  See~\cite{cade/UrbanSPV08,2011arXiv1108.3446A} for the details of how such
machine learning is organized in the context of a large deductive
repository like MML, and for detailed statistics on how it improves
existing premise selection methods.  
This service is very fast, taking typically less than a second.  The
hints are again HTML-ized and
inserted into an explanation box, as shown in
Figure~\ref{Hints}.  A similar hint function is accessible also from the
Emacs mode.

\vspace{-2em}

\section{Initial evaluation of the ATP-supported  authoring}
\label{sect:experience}

In January 2011
the second author was developing a new formalization of simple graphs
as 1-dimensional simplicial complexes. The goal of the article was to develop enough theory to
prove the construction of the Mycielskian of a graph.  This
construction was used by Mycielski~\cite{Mycielski55} to prove
existence of triangle-free graphs with arbitrarily large chromatic
number.

\vspace{-1em}

\subsection{Axiom debugging}
The \MizAR service was used on a \mizar article that was in an
early stage of development, such that even the formalization of basic
definitions was still not ironed out. Only very basic notions of set 
theory are used in the article: empty set, singleton, ordered and
unordered pairs, union of a family of sets, subset, finite set,
cardinality of a finite set, partition of a set, Cartesian product,
basic properties of relations and functions, and not much more.  It was
of interest to see how the \MizAR service performs under such conditions,
since the \mizar library contains thousands of facts using only these
basic notions.

An initial surprise came when the ATP was able to prove almost anything. It turned out that
the following statement:

\begin{small}
\begin{verbatim}
   theorem SG1:
   for G being SimpleGraph holds {} in G @proof end;
\end{verbatim}
\end{small}
was to blame.\footnote{The \texttt{@proof} syntax tells \mizar to skip checking of such proof block.} 
While developing a new formalization we frequently state
similar, simple facts in a top-down manner, and leave them unproven while focusing on more interesting
pieces. We had carelessly stated that \verb+{}+ is in every simple
graph, even in an empty graph.  With an easy contradiction derivable
from the axioms, a (refutational) ATP can justify anything.
 After correcting this to:

\begin{small}
\begin{verbatim}
   theorem SG1:
   for G being non empty SimpleGraph holds {} in G @proof end;
\end{verbatim}
\end{small}
the ATP was still surprisingly successful. A similar unproven statement
\begin{small}
\begin{verbatim}
  theorem SG0:
  for G being SimpleGraph holds G = { {} } \/ Vertices G \/ Edges G @proof end;
\end{verbatim}
\end{small}
turns out to be false, as it fails when {\tt G} is empty.  The
presence of this unproven fact allowed the ATP to prove many other facts
in a rather unexpected way. This was again corrected by requiring
{\tt G} in {\tt SG0} to be nonempty and later led us to revise the definition
of {\tt SimpleGraph}.
Thus, the ATP helped to straighten out the basic
definitions before we did more proofs that we deemed interesting, but which
would have been based on unproven, contradictory lemmas about empty
graphs. 

\vspace{-1em}

\subsection{Deciphering ATP proofs}
There are times when ATP manages to find a proof for a fact that
is worth including in the \mml as an exportable (reusable) item.
Such items are marked {\tt theorem} in \mizar.  Here is an
example

\begin{small}
\begin{verbatim}
   theorem Aux1a:
   for x, X being set holds not [x,X] in X
\end{verbatim}
\end{small}

for which the ATP returns the following list of premises:

\begin{small}
\begin{verbatim}
   by ENUMSET1:69,ZFMISC_1:12,ZFMISC_1:38,TARSKI:def 5,ORDINAL1:3;
\end{verbatim}
\end{small}

Even though this is very far from deep ATP proofs,
the example shows that it may be a bit of a challenge to
convert the resolution proof found by ATP into a 
sequence of inference steps that are understandable to humans and
acceptable by the \mizar verifier.
After examining the ATP proof we constructed a detailed
justification by hand, using the {\tt proof} construct.

\begin{small}
\begin{verbatim}
   theorem Aux1a:
   for x, X being set holds not [x,X] in X
   proof
     let x, X be set such that
   A: [x,X] in X;
   B: [x,X] = { {x,X}, {x} } by TARSKI:def 5;
   C: {x,X} in { {x,X}, {x} } by ZFMISC_1:38;
   D: X in {x,X} by ZFMISC_1:38;
     thus contradiction by A, B, C, D, ORDINAL1:3;
   end;
\end{verbatim}
\end{small}

Later a more natural proof of this little fact was found, directly using 
the definition of an unordered pair.
Directing the ATP to use premises prefered by the user would be
an interesting future research.

\subsection{Overall ATP efficiency and experience}

Of the few hundred non-trivial inferences that were tried in the {\tt
  SCMYCIEL} article, ATP managed to solve around 40\%, which,
is surprisingly close to the success rate of Sledgehammer on
non-trivial goals~\cite{sledgehammer10}.  On the other hand, ATP 
can re-prove 86\% of the inferences if it is told which premises
were used by humans.  This means that more precise narrowing of
potential premises is a vital issue for the ATP service, which could
particularly benefit from learning from the large number of previous
proofs~\cite{2011arXiv1108.3446A}.  As mentioned in Section~\ref{ATPCalls}, an earlier
experiment using several ATP systems 
has shown that with smarter premise selection more than 99\% of atomic inferences can be 
re-proved~\cite{mics/UrbanS08}.\footnote{Precisely, 99.8\% atomic inferences (6751 out of 6765) were re-proved 
automatically. The 14 remaining problems were also proved by ATPs after manual premise selection.}
The interactive ATP service helped in several ways:
\begin{itemize}
\item ATP managed to directly prove some lemmas that require a structured proof for the \mizar
  verifier. This is not a big surprise, as the \mizar verifier uses only
  pattern matching and very limited resolution. The feedback from the ATP
  system was quite helpful, as it is much easier to write a detailed proof
  when one knows the facts that suffice for the proof.
\item ATP turned out to be a search tool in a rather unexpected
  way. More than once the ATP system indicated that a local lemma had been 
  formed, from which the given formula followed in one step, while we
  were about to write several inference steps.
\item ATP systems found proofs quite different from what
  the user had in mind.  Sometimes it found large sets of premises when some
  small collection of premises sufficed. The converse also happened.
\item When the ATP system finds a proof, it returns all \mizar items
  that it used.  The feedback
  also includes those \mizar items that are tacitly processed by
  the verifier, and they cannot be referenced in a \mizar text.  
  This feedback information led us to a better understanding
  of the task at hand.
\end{itemize}

\section{Conclusions, Related and Future Work }
\label{Future}

The \MizAR service allows authors 
to use a number of auxiliary tools on \Mizar articles.
The use-cases range from using HTMLization to
disambiguate complicated \Mizar texts, using ATP systems to find new
proofs, explaining \Mizar inferences, and finding counterexamples, to
using AI-based techniques for proof advice. The system features both a
web and an Emacs interface, allowing flexible switching between
\emph{reading and exploration} mode and \emph{authoring} mode.

Related work goes back at least to Dahn's~\cite{DahnW97} work in 90's
on ILF and its \Mizar-to-ATP bridge, Harrison's and Hurd's work on
Meson~\cite{Har96} and Metis~\cite{hurd2003d} used in HOL (Light), and the recent work by
Paulson et al.~\cite{sledgehammer10} on linking Isabelle/HOL
with ATP systems. 
A detailed comparison of systems bridging ITP with ATP systems 
is beyond the scope of this paper.

There are many directions for future work in this setting, some of them
mentioned above.
Several versions of the MML are now present on the servers in 
text, HTML, MPTP, and TPTP formats, but are not directly editable by the users.
Giving the user the ability to edit the supporting \MML leads in the 
direction of formal mathematical wikis, with all the interesting persistence, 
versioning, linking, user-authentication, and dependency problems to solve.
Merging the current \Mizar wiki  development with the services presented
here is obvious future work. This should form a rich collaborative platform for formal mathematics, with a large
number of services providing strong automated support, and exposing
the functionalities that make formal mathematics so interesting.
We foresee a large amount of work on making the system stronger, 
more attractive and responsive.
\begin{itemize}
\item 
  We would like to speed up all \MizAR services through more parallel processing 
  once the necessary hardware is available.
\item There seems to be no end to improving techniques
for hint selection in large libraries.  We consider such techniques
crucial to the success of bridging ITP and ATP systems.
\item There is an urgent need for converting the very verbose
  (typically refutational) proofs found by ATP systems into structured
  and simple to check proofs.\footnote{We
  consider it a good feature that \mizar does not allow complicated,
  fragile, and slow proof finding procedures as a part of the core
  proof checking. The twenty years of experience with daily
  large-scale theory refactoring of \MML has taught the \mizar
  community that such fragility and slowness should be avoided.  We
  strongly believe that the way from automatically found proofs
  to proofs in the \MML leads through suitable semi-automated refactoring into
  structured proofs which are perceived as obvious by humans.  
A related recent
  effort is described in~\cite{VyskocilSU10}. }
  As ATP systems are getting stronger and
  more useful for finding proofs, this problem is becoming more
  pressing.
\end{itemize}

Our initial experience with the interactive ATP service for \mizar
authors is encouraging.  Despite very large library context, we get
decent automated help both in finding justification for proof steps,
as well as in `debugging' the conceptual framework of a new
formalization.

\bibliographystyle{plain}
\bibliography{Bibliography,escape}
\end{document}